\documentclass[a4paper,aps,superscriptaddress,floatfix,nofootinbib,twocolumn,bibtex]{revtex4}

\usepackage{bbold}
\usepackage{bbm}
\usepackage[pdftex]{graphicx}
\usepackage{latexsym,amsmath,verbatim}
\usepackage{color}
\usepackage{rotating}
\usepackage{multirow}
\usepackage[english]{babel}
\usepackage{sidecap}

\begin{document}

\title{Driving interconnected networks to supercriticality}

\author{Filippo Radicchi}
\affiliation{Center for Complex Networks and Systems Research, School of Informatics and Computing, Indiana University, Bloomington, USA}
\email{filiradi@indiana.edu.}

\begin{abstract}
\noindent Networks in the real world do not exist as isolated entities, 
but they are often part of more complicated structures composed
of many interconnected network layers. 
Recent studies have shown that such mutual dependence 
makes real networked systems potentially exposed to atypical 
structural and dynamical behaviors, and thus there is a urgent necessity to 
better understand the mechanisms at the basis of these
anomalies. Previous research has mainly focused on the 
emergence of atypical properties in relation with
the moments of the intra- and inter-layer degree
distributions. 
In this paper, we show that an additional ingredient plays
a fundamental role for the possible scenario that
an interconnected network can face: the correlation between
intra- and inter-layer degrees. For sufficiently
high amounts of correlation, an interconnected
network can be tuned, by varying
the moments of the intra- and inter-layer degree distributions, in distinct 
topological and dynamical regimes. When instead the
correlation between intra- and inter-layer degrees is
lower than a critical value, the system enters in a supercricritical
regime where dynamical and topological phases are not
longer distinguishable.
\end{abstract}

\maketitle

\noindent The traditional study of networks
as isolated entities has been recently
overcome by a more realistic approach 
that accounts for interactions between networks~\cite{Buldy10}.
Networks in the real world are in fact
often, if not always, mutually connected: 
social networks (e.g., Facebook, Twitter) are coupled because
they share the same actors~\cite{szell10};
transportation networks  are composed of
different layers (e.g., buses, airplanes) with common nodes standing
for the same geographic locations~\cite{Barthelemy11}; the functioning
of communication and power grid systems depend one on the
other~\cite{Buldy10}. 
As the properties of an isolated network are 
not trivially deducible 
from those of the individual vertices that
are part of it, at the same strength
decomposing an interconnected network and studying each component in
isolation does not allow to understand the whole system and its
dynamics. 
Indeed, interconnected networks often exhibit properties that
strongly differ form those typical of isolated
networks: for example, 
structural~\cite{Buldy10} and dynamical~\cite{Radicchi13}
transitions, that are usually continuous
in isolated networks~\cite{Dorogovtsev08}, 
may become discontinuous in interconnected networks.
So far, theoretical studies have pointed out that
phase transitions in random interconnected networks
become abrupt only if the strength (or density)
of the interconnections is sufficiently 
large when compared to the first
moment of the strength (or degree) distribution
of the whole network~\cite{Parshani10,Gao11, Gao12, Seung12, Radicchi13}.
In this paper, we show that 
the situation is not so simple, but
depending on the combination of basic
structural features -- strength of interconnections, 
first two moments of the degree  distribution of
the entire network, and  correlation between
intra- and inter-layer degrees -- 
an interconnected network explores a complicated scenario where its critical
properties can drastically mutate.

\

\noindent In the following, we focus our attention on the case of 
two arbitrarily interconnected
network layers, and study the spectral properties
of its associated normalized laplacian~\cite{Chung_book}. 
The choice of this matrix is not
arbitrary. The normalized laplacian
of a network is in fact an object of fundamental importance
for the understanding of its structural and 
dynamical properties, sometimes even more
than the adjacency matrix. For example, the spectrum of the 
normalized laplacian is used in spectral
graph clustering to determine the internal
organization of a graph~\cite{Shi97,Rosvall08}, and
many useful measures, such as 
graph energy~\cite{Cavers10}, graph conductance and 
resistance~\cite{Doyle84},
and the Randi\'c index~\cite{Klein93}, are quantifiable in terms of
the eigenvalues of the normalized laplacian.
Also, the normalized laplacian
fully describes the behavior of one of the most important 
dynamical processes studied in science, from
biology~\cite{Berg93} to computer science~\cite{Brin98}, from
ecology~\cite{Codling2008} to finance~\cite{Burton73} and
physics~\cite{Weiss94}: random walk dynamics. 
The normalized
laplacian is representative for both the classical~\cite{Chung_book} 
and the quantum~\cite{Faccin13} versions of random walk
in networks, and more in general for every time reversible
Markov chain~\cite{Lovasz1996}.

\begin{figure*}[!htb]
\begin{center}
\includegraphics[width=0.75\textwidth]{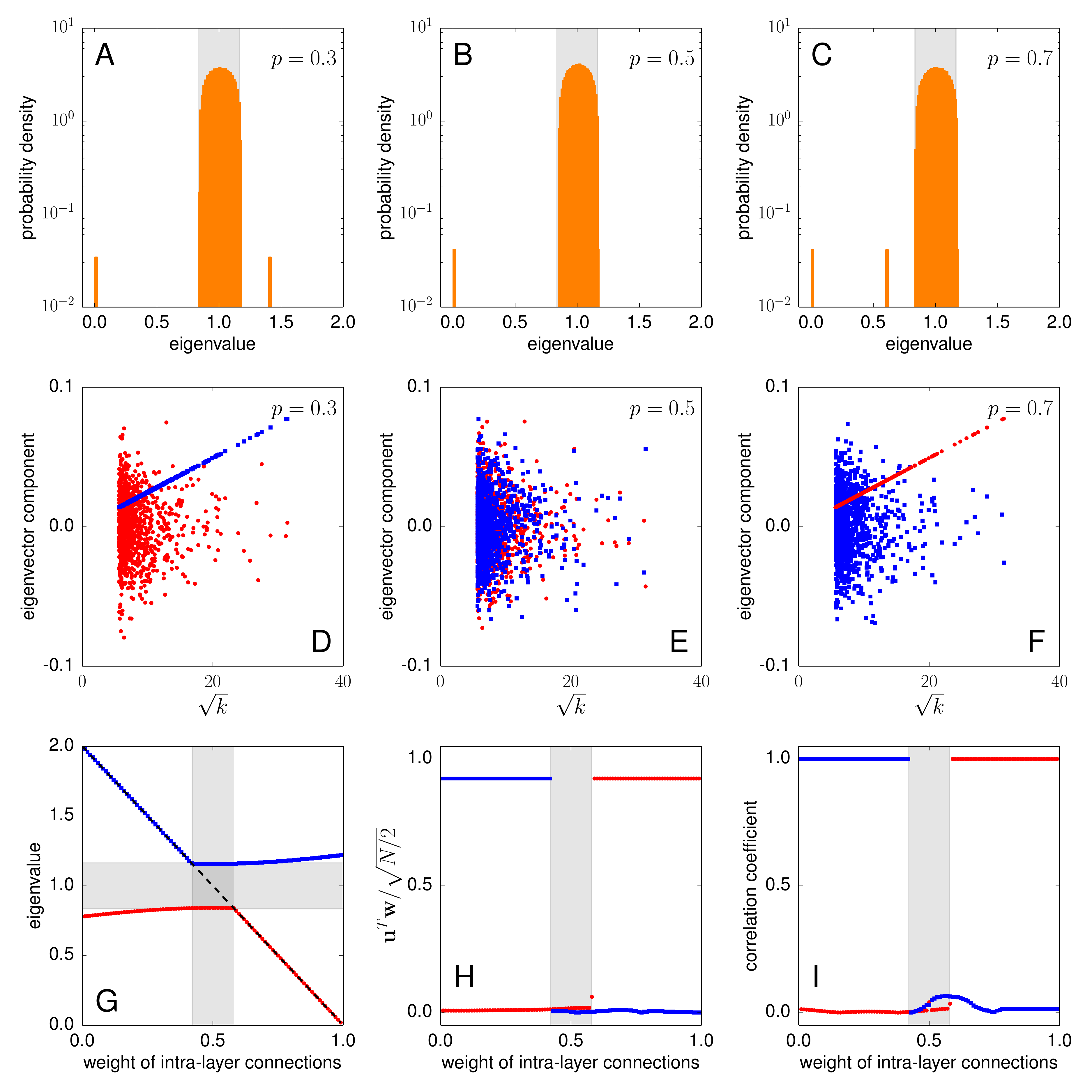}
\end{center}
\caption{
{\bf Abrupt transition in interconnected networks}.
Spectral analysis of the normalized laplacian $\mathcal{L}$
of an interconnected network composed of two network layers 
of size $N=1024$. The degree sequence
is composed of integers extracted from a power-law
probability distribution with 
exponent $\gamma=2.5$ defined on the
support $[32,N]$. 
Intra- and inter-layer degree sequences
are perfectly aligned
so that the correlation term 
reads $\xi = \sum_{k^{in},k^{out}} k^{in}\, k^{out}\; P(k^{in}, k^{out})= \langle k^2 \rangle$. 
As the weight of the intra-layer
layer connections $p$ varies, the system explores
different regimes clearly visible from the
spectrum of $\mathcal{L}$.
{\bf A} For $p=0.3$, only the largest eigenvalue
$\nu_{2N}$ is well separated from the rest of the spectrum.
The system is said to be in the ``bipartite'' regime ($\mathcal{B}$-phase).
{\bf B} For $p=0.5$, there are no detached eigenvalues.
The system is in the ``indistinguishable'' regime ($\mathcal{I}$-phase).
{\bf C} For $p=0.7$, the second smallest eigenvalue $\nu_2$ is well 
separated for the rest of the spectrum.
The system is in the ``decoupled'' regime ($\mathcal{D}$-phase).
{\bf D} In the $\mathcal{B}$-phase, the components
of largest eigenvector $\mathbf{w}_{2N}$ 
corresponding to a single network layer (blue squares)
are linearly correlated with the square root of the
respective node degrees $\sqrt{k}$, while
those of the second smallest
eigenvector  $\mathbf{w}_{2}$ (red circles) are not.
{\bf E} In the $\mathcal{I}$-phase, there is no
correlation between the components of
$\mathbf{w}_{2N}$ ($\mathbf{w}_{2}$) and $\sqrt{k}$.
{\bf F} In the $\mathcal{D}$-phase, the components
of second smallest eigenvector $\mathbf{w}_{2}$ 
are linearly correlated with $\sqrt{k}$, while the components of  
$\mathbf{w}_{2N}$ are not.
{\bf G} The transition points 
$p_c^- \simeq 0.42$ and $p_c^+ \simeq 0.58$ 
(delimiting the vertical gray area)
between the different regimes
correspond to the points in which our 
prediction [Eq.~(\ref{eq:pc_thre}), black dashed line]
enters the continuous band of the spectrum [Eq.~(\ref{eq:chung}), 
horizontal gray area].
Red circles refer to $\nu_2$, while blue squares
refer to $\nu_{2N}$.
{\bf H} The sums of the components of 
$\mathbf{w}_2$ (red circles) and 
$\mathbf{w}_{2N}$ (blue squares)
corresponding to nodes of a single
network layer can be used as a order parameter to monitor
the transition between the different regimes ($\mathbf{u}$ is the vector whose components are all
equal to one).
{\bf I} As noted above, in the $\mathcal{D}$-phase ($\mathcal{B}$-phase), the
components of $\mathbf{w}_2$ ($\mathbf{w}_{2N}$)
are perfectly correlated with the square 
root of the node degrees.
}
\end{figure*}

\noindent Let us consider a symmetric and 
weighted interconnected network formed by two interconnected 
layers of identical size $N$ whose adjacency matrix $G$ 
is written in the block form
\begin{equation}
G = p \, G_{in} + (1-p) \, G_{out}  = p 
\left(
\begin{array}{cc}
A & \emptyset
\\ 
\emptyset & B
\end{array}
\right)
+
(1-p)
\left(
\begin{array}{cc}
\emptyset & C
\\ 
C^T & \emptyset
\end{array}
\right) \; .
\label{eq:adj_sp}
\end{equation}
$A$, $B$ and $C$ are
$N \times N$ square matrices.
$A$ and $B$ are symmetric matrices, 
while $C$ is not necessarily symmetric. 
According to Eq.~(\ref{eq:adj_sp}), the adjacency matrix $G$ is a
convex linear combination of the intra- and inter-layer
adjacency matrices $G_{in}$ and $G_{out}$, 
and the parameter $p \in [0,1]$ 
serves to continuously tune the entire network from 
a bipartite graph ($p=0$) to
a perfectly decoupled networked system ($p=1$).
The normalized graph laplacian associated
with the adjacency matrix of Eq.~(\ref{eq:adj_sp}) is defined as
\begin{equation}
\mathcal{L} = \mathbbm{1} - D^{-1/2} G D^{-1/2} \; .
\label{eq:lap}
\end{equation}
In the definition of $\mathcal{L}$, $\mathbbm{1}$ is the
identity matrix, and $D^{-1/2}$ is a diagonal matrix whose diagonal
elements are equal to the inverse of the square root 
of the node strengths~\cite{Chung_book}.
Since $G$ is symmetric, 
all the eigenvalues of $\mathcal{L}$
are real numbers in the range $[0,2]$. For simplicity,
we sort them in ascending order such that
$0 = \nu_1 \leq \nu_2 \leq \cdots \nu_{2N} \leq 2$,
and we indicate with $\mathbf{w}_i$ the eigenvector
associated to $\nu_i$. The smallest eigenpair 
$(\nu_1, \mathbf{w}_1)$ is generally said to be 
``trivial'' because it depends only on the
strength sequence of the graph.
We always have $\nu_1=0$ and
$\mathbf{w}_1^T = 1/\sqrt{2 N \langle s \rangle} \left(\sqrt{s_1}, 
\ldots, \sqrt{s_{2N}}\right)$, with $\langle s \rangle$ first
moment of the strength distribution, and
$s_i$ strength of the node $i$. 
When translated into the language of random walk 
dynamics, $\mathbf{w}_1$ tells us that, 
independently of the initial conditions and the topology
of the network, the stationary probability to find
the random walker on a given node is
linearly proportional to its strength.
The other two external eigenvalues $\nu_2$
and $\nu_{2N}$, and their associated eigenvectors
$\mathbf{w}_2$ and $\mathbf{w}_{2N}$, are
much more meaningful from the structural and 
dynamical points of view. $\nu_2$ is always strictly
larger than zero in graphs composed
of a single connected component. 
The associated 
eigenvector $\mathbf{w}_2$
is typically used in graph clustering
to determine the bisection corresponding to
the minimum of the normalized
cut of the graph~\cite{Shi97}. 
In terms of random walk dynamics, such
eigenvector is also very meaningful because the signs 
of its components define the so-called
almost invariant sets of the
graph~\cite{Schwartz06,Rosvall08}. 
In essence, a random walker spends most 
its time moving among nodes whose
correspondent components in $\mathbf{w}_2$ have the
same sign, and less frequently jumps between 
nodes whose correspondent components in $\mathbf{w}_2$ have
different signs. 
The largest eigenvalue $\nu_{2N}=2$
only if the graph is bipartite, while 
$\nu_{2N} < 2$ in all other cases, and the components
of $\mathbf{w}_{2N}$ identify the
bipartite components of the graph.
A random walker typically jumps between pairs
of nodes that correspond to components 
of $\mathbf{w}_{2N}$ with different signs.

\

\noindent To understand the spectral properties of
$\mathcal{L}$, 
we restrict our attention to the
case of random network
models generated according to the
following procedure. Intra- and inter-layer degrees 
of both network  layers $A$ and $B$ are respectively given by
the same identical intra- and inter-layer degree sequences 
$k_{in} = \left\{k_1^{in}, k_2^{in}, \ldots, k_N^{in}\right\}$
and
$k_{out} = \left\{ k_1^{out}, k_2^{out}, \ldots, k_N^{in} \right\}$.
The numbers appearing in the two degree sequences
are also identical except for their order of appearance.
The intra- and inter-layer degree distributions 
are thus identical $P(k^{in}) = P(k^{out}) = P(k)$, while the
alignment between the two sequences regulates the joint probability 
distribution $P(k^{in}, k^{out})$ of both
network layers. In the construction
of the network,
internal and external connections are randomly placed with
the only constraints of preserving the {\it a priori} provided
degree sequences and generating
network layers composed of a single connected component, so that
$\nu_2>0$ and $\nu_{2N}<2$ for every $p \in (0,1)$.
It is worth to briefly discuss the meaning, limitations
and justifications of the previous assumptions. 
The fact that intra- and inter-layer connections 
are randomly placed allows us to say 
that no substructure is present
in layers $A$ and $B$. Although this is
not a good assumption for real networks that instead
often exhibit modular structure
at single layer level~\cite{Fortunato10}, it is
a necessary assumption to concentrate our attention
only on the effect of the layer structure on the 
spectrum of $\mathcal{L}$. 
The fact that the intra- and inter-layer degree distributions are
identical, and both network layers
have the same degree distribution and 
joint probability function $P(k^{in}, k^{out})$
are also non realistic assumptions. 
In real world interconnected networks, in fact, one should expect 
the various distributions to be different. 
Although our general solution requires a much weaker assumption (see SI),
here we decided to impose these constraints for illustrative purposes. 
The scenario is in fact already so 
much rich that having an additional symmetry
in the system makes easier the
explanation of the various behaviors that
the system can exhibits. 

\

\noindent If we think in terms of random walk dynamics, 
our network models are symmetric in two ways. First, 
the stationary probability to find
the random walker in one of the two layers
is equal to $1/2$. Second, they introduce a symmetry
around the point $p=1/2$ for the various
regimes in which the system
can be tuned by varying $p$. Intuitively, we
expect the following: (i) For $p < 1/2$, the network should
be in a ``bipartite'' regime ($\mathcal{B}$-phase) with the random walker
moving more often from one layer to the other, and
less frequently diffusing in the same layer;
(ii) For $p > 1/2$, the network should be in a ``decoupled''
regime ($\mathcal{D}$-phase) with the random walker more likely diffusing
between nodes of the same layer, and less likely jumping
between layers;  (iii) For $p \simeq 1/2$, the network should be in a
``indistinguishable'' regime ($\mathcal{I}$-phase), with the random
walker moving between and among layers with
equal probability. Please note that, in the $\mathcal{I}$-phase, the system
is topologically and dynamically identical to
a single-layer network, thus in a potentially safe state.
On contrary, when layers are structurally and dynamically 
distinguishable (i.e., $\mathcal{B}$- and $\mathcal{D}$-phases), 
then the network behaves as 
an interconnected network, and, as such, is
exposed to potentially catastrophic failures
or anomalous dynamical behavior.
Intuitively, one should expect
that the network is driven in a continuous way among
the different phases as $p$ is varied.
In reality, the scenario 
is much more complicated and appealing.

\begin{figure}[!htb]
\begin{center}
\includegraphics[width=0.45\textwidth]{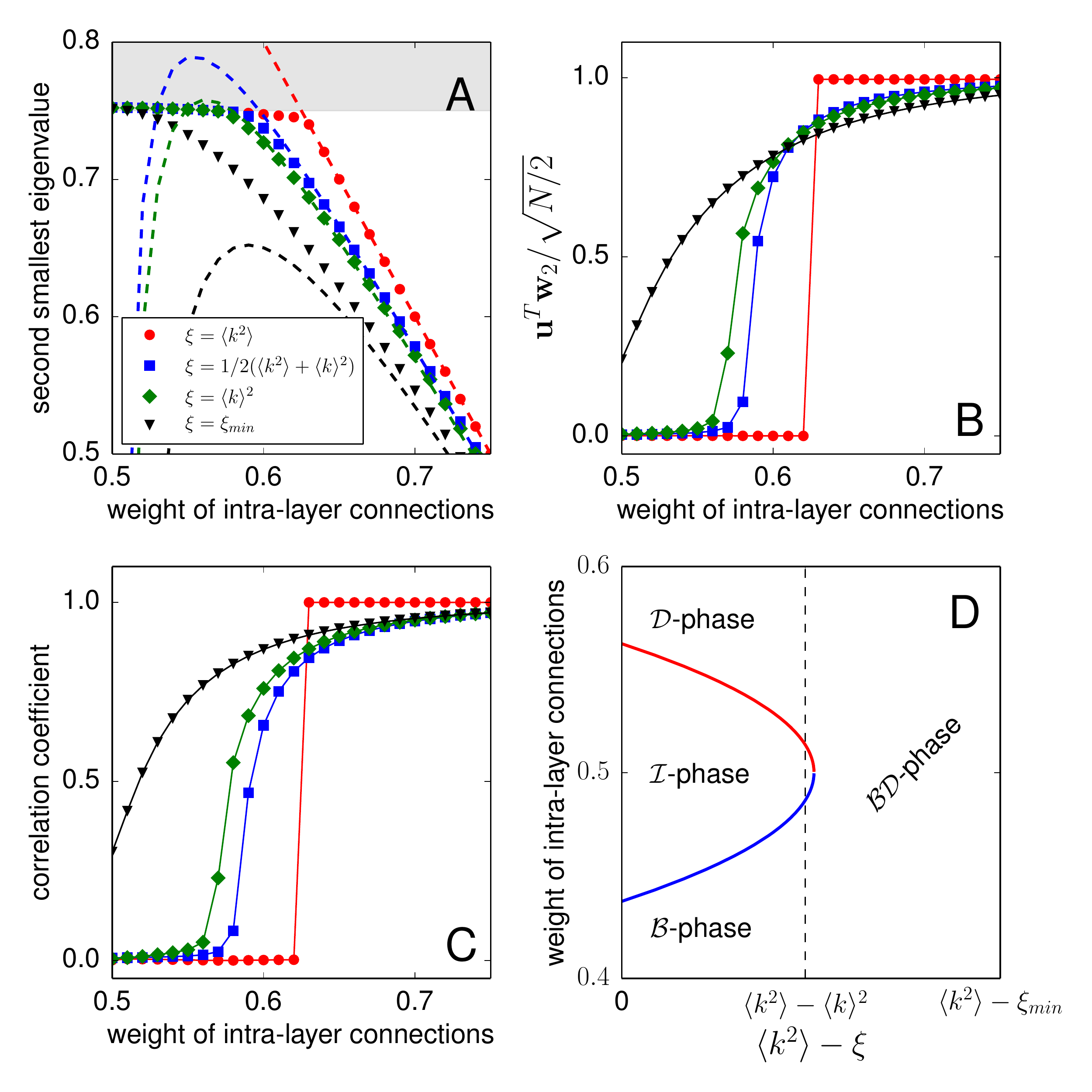}
\end{center}
\caption{
{\bf Phase diagram}.
We consider interconnected networks
composed of two layers 
of size $N=8192$ with intra- and inter-layer degrees obeying a 
Poisson distribution with average $\langle k \rangle =32$. 
{\bf A} Second smallest eigenvalue $\nu_{2}$
as a function of the weight
of the intra-layer connections $p$. 
For values of the correlation term $\xi$ in the range
$[\langle k \rangle^2, \langle k^2 \rangle]$,  
the results of numerical computations 
(symbols) are in very good agreement with our 
analytic predictions (dashed lines). The critical
point $p_c^+$ is determined as the point for which
the dashed lines enter the spectral band (gray area).
For $\xi=\xi_{min}$, our predictions fail 
to correctly describe how $\nu_2$ changes 
as a function of $p$. {\bf B} As in Fig.~1, we monitor
the transition between the
$\mathcal{D}$- and the $\mathcal{I}$-phase, 
by looking at the order parameter
$\mathbf{u}^T\mathbf{w}_2/\sqrt{N/2}$.
As $\xi$ decreases, the change of the order
parameter becomes smoother, and
the value of the critical
point gets closer to $p=0.5$. For
$\xi \in [\langle k \rangle^2, \langle k^2 \rangle]$,
the order parameter
goes to zero if $p<p_c^{+}$.
For $\xi=\xi_{min}$ instead, the order
parameter is larger than zero even for $p=0.5$. 
{\bf C} The same behavior described
for the order parameter is visible if
we monitor the transition through the
linear correlation coefficient between the eigenvector
components and the r.h.s of Eq.~(\ref{eq:ans}).
{\bf D} Phase diagram representing the
expected behavior of the system in the thermodynamic 
limit. The red line stands for the critical 
value $p_c^+$, while the blue
line represents $p_c^-$ [Eq.~(\ref{eq:thre})].
These critical values are plotted as functions
of the quantity $\langle k^2\rangle - \xi$, 
ranging between zero and
$\langle k^2\rangle - \xi_{min}$. As a reference, 
the dashed line represents the point at which
$\xi=\langle k^2 \rangle$.
Although the curves corresponding
to $p_c^+$ and $p_c^-$ do not intersect, for values of the
correlation term $\xi$ sufficiently low, their analytic
continuations suggest the disappearance
of the $\mathcal{I}$-phase,  
and the appearance of a supercritical regime
where both the bipartite and the
decoupled phases are present (i.e., $\mathcal{BD}$-phase).
}
\end{figure}

\noindent Let us first focus on the exactly solvable case
in which the intra- and inter-layer degree sequences of both layers 
are ordered in the same way ($k^{in}_i = k^{out}_i =k_i$, for all nodes $i$). 
Please note that in this case the node strengths equal
the node degrees, and thus they do not
change with $p$ [$s_i = p k_i^{in} + (1-p) k_i^{out} =k_i$, for
all nodes $i$]. In addition to the trivial eigenpair,
$\mathcal{L}$ has always another eigenpair 
$(\nu^*, \mathbf{w}^*)$ given by 
\begin{equation}
 \nu^* = 2(1-p)  \qquad \textrm{ and } \qquad w^*_i = c_i \, \sqrt{k_i}  \quad , \, \forall \, i =1 ,\ldots, 2N \; ,
\label{eq:exact}
\end{equation}
where $c_i = c$, for $i=1, \ldots, N$,
and $c_i = - c$, for $i=N+1, \ldots, 2N$, and
$c=1/\sqrt{2 N \langle k \rangle}$ 
with $\langle k \rangle$ 
first moment of the degree distribution (see SI).
Eq.~(\ref{eq:exact}) tells us that
the eigenvector $\mathbf{w}^*$ is able
to perfectly reflect the layer structure of the system. 
Such eigenvector is associated to an eigenvalue
that spans the entire range $[0,2]$, and therefore 
must intersect all the other eigenvalues of $\mathcal{L}$. 
At each point of intersection, two orthogonal eigenvectors
correspond to the same eigenvalue, and thus, in physical
terms, each point of intersection
corresponds to a degenerate energy level. 
Such phenomenology is typical
of discontinuous phase
transitions~\cite{Blundell06}. Please note
that this happens in networks
of any size, and for any degree distribution.
The changes between
the various phases that we described
above are of the same nature: as $p$ is 
increased, the system moves discontinuously
from the $\mathcal{B}$-phase to the $\mathcal{I}$-phase, 
and from the
$\mathcal{I}$-phase to the $\mathcal{D}$-phase.
The $\mathcal{B}$-phase can be identified as
the regime in which $\nu^*$ corresponds to the largest
eigenvalue of $\mathcal{L}$. Similarly,
the $\mathcal{D}$-phase can be determined as the
regime in which $\nu^*$ represents the
second smallest eigenvalue of $\mathcal{L}$. 
Finally in the $\mathcal{I}$-phase, $\nu^*$
is inside the spectrum of $\mathcal{L}$.
The values of $p$ for which
phase transitions occur
can be approximately estimated with the following
argument. Numerical computations of the 
spectrum of $\mathcal{L}$ show
that, whereas the eigenvalue $\nu^*$ is rapidly changing
as a function of $p$, all other eigenvalues vary much 
more slowly (see upper panels of Fig.~1). 
The graph obtained for $p=1/2$ can be thus be used as a representative for
this almost constant behavior. We can then 
approximate, thanks to the prediction of Chung {\it et al.}~\cite{Chung03}, 
the expected spectral radius 
of the normalized laplacian of a random graph 
with prescribed degree distribution as
\begin{equation}
\max_{i > 1} \left| 1 - \nu_i \right| \simeq \frac{2}{\sqrt{2 \langle k \rangle}}\; .
\label{eq:chung}
\end{equation}
Eq.~(\ref{eq:chung}) generally provides very
good estimates for
the spectral radius of $\mathcal{L}$ for 
random networks with poissonian and power-law degree distributions.
By comparing $\nu^*$ to the r.h.s. of Eq.~(\ref{eq:chung}), we finally
find that the two transition points are
\begin{equation}
p_c^{+,-} \simeq \frac{1}{2} \pm \frac{1}{\sqrt{2 \langle k \rangle}} \; .
\label{eq:pc_thre}
\end{equation}
In Eq.~(\ref{eq:pc_thre}), the solution with 
the minus sign corresponds to critical point of
the transition between the $\mathcal{B}$- and 
$\mathcal{I}$-phases, while the solution with the plus sign corresponds to
threshold between the $\mathcal{I}$- and 
$\mathcal{D}$-phases. As expected,
such predictions are in very good agreement
with the results of numerical experiments (see Fig.~1).

\

\noindent Next, we relax the former assumption about
perfectly aligned degree sequences, and we let 
$k_{out}$ correspond to a slightly modified permutation
of the sequence $k_{in}$. Unfortunately, our
theory works only for Poisson degree distributions, but
the results we obtain allows us to
understand also the qualitative behavior of
networks with power-law degree distributions.
If degrees are random variates extracted
from a Poisson distribution,
we are able to numerically
show that the $i$-th component of the second smallest (only in 
the $\mathcal{D}$-phase)
or the largest eigenvector (only in the $\mathcal{B}$-phase) $\mathbf{w}^*$
obeys the relation
\begin{equation}
w^*_i \; \simeq  c_i \; \frac{p \,k_i^{in} - (1-p) \, k_i^{out} }{ \sqrt{p \,k_i^{in}  + (1-p) \, k_i^{out}}}  \; ,
\label{eq:ans}
\end{equation}
where $c_i = c$, for $i=1, \ldots, N$,
and $c_i = - c$, for $i=N+1, \ldots, 2N$, and
$c$ is a proper normalization constant.
This essentially means that the $i$-th component 
of the eigenvector $\mathbf{w}^*$ is proportional to
the difference between the internal and external strengths
divided by the square root of the total strength
of node $i$ (Fig.~S1). The validity of 
Eq.~(\ref{eq:ans}) is testified by the
very high linear correlation coefficients
that one can measure between the components
of the eigenvectors (numerically computed),
and the r.h.s. of Eq.~(\ref{eq:ans}).
Using Eq.~(\ref{eq:ans}) as ansatz
for the solution of the eigenvalue
problem, we are able to determine that the
eigenvalue of $\mathcal{L}$ 
associated to the eigenvector $\mathbf{w}^*$ 
of Eq.~(\ref{eq:ans}) is given by
\begin{equation}
\nu^*  \simeq 1 - \frac{\left(2 p^2 -2p +1 \right) \, \langle k^2 \rangle  - 2 p (1-p) \, \xi }{\left(2p -1 \right) \, \langle k^2 \rangle} \; ,
\label{eq:nu}
\end{equation}
and thus  by the ratio between a quadratic
and a linear function of $p$. 
The function appearing at
the denominator depends only on the second moment $\langle k^2\rangle$
of the degree distribution.
The function appearing at the numerator, instead,
depends on the second moment
of the degree distribution, and the
correlation between intra- and inter-layer degrees 
\begin{equation}
\xi  = \sum_{k^{in},k^{out}} k^{in}\, k^{out}\; P(k^{in}, k^{out}) \; .
\label{eq:corr}
\end{equation}
By comparing Eq.~(\ref{eq:nu}) with the expected 
spectral radius of Eq.~(\ref{eq:chung}), we finally arrive
to the prediction of the critical points
\begin{equation}
p_c^{+,-} \simeq \frac{1}{2} \left[ 1 \pm \frac{\frac{2}{\sqrt{2 \langle k \rangle}} \langle k^2 \rangle + \sqrt{\xi^2 + \langle k^2 \rangle^2 \left(\frac{2}{\langle k \rangle} -1\right) } }{2\left( \langle k^2 \rangle + \xi \right)} \right] \; .
\label{eq:thre}
\end{equation}
When the term inside 
the square root of Eq.~(\ref{eq:thre}) is larger than
zero, we have an intersection between
$\nu^*$ and the
rest of spectrum of $\mathcal{L}$. This indicates
the presence of an abrupt transition between the
corresponding phases. Strictly speaking, these 
eigenvalues do not intersect in finite size random networks,
as predicted by the
Wigner-von Neumann noncrossing rule~\cite{vonNeumann29,Lax07}, 
however, the abrupt nature
of the transition appears clear already for
networks of moderately large size.
When the term in the square root of Eq.~(\ref{eq:thre}) is equal to zero,
the non trivial
eigenvalue $\nu^*$ touches the
rest of spectrum of $\mathcal{L}$ tangentially at $p_c$. 
In this case, the corresponding transition becomes continuous.
The non existence
of a real solution in Eq.~(\ref{eq:thre}) reveals
instead the absence of a crossing between $\nu^*$
and the rest of the spectrum even in infinite size networks,
and thus is the sign of a dramatic change
in the behavior of the system.

\

\noindent 
Eqs.~(\ref{eq:nu}) and~(\ref{eq:thre}) show that the
the entire scenario
is characterized by two main factors. On one end, 
the degree distribution, with its 
first two moments, play a fundamental 
role for the determination
of the various phases. On the other hand, 
the degree distribution alone is not able to explain the
behavior of the system. At parity
of moments in fact, the features of the system 
can be still drastically changed by the correlation term 
defined in Eq.~(\ref{eq:corr}).
To better understand how the various possibilities are 
regulated by the interplay between the
correlation term $\xi$ and the moments of the
degree distribution,
we  vary the correlation term by regulating the
alignment between the intra- and inter-layer degree
sequences $k_{in}$ and $k_{out}$. 
If the order of the sequences 
$k_{in}$ and $k_{out}$ is identical,
$\xi=\langle k^2 \rangle$,  and 
Eqs.~(\ref{eq:nu}) and~(\ref{eq:thre})
correctly reduce to Eqs.~(\ref{eq:exact}) and~(\ref{eq:pc_thre}), 
respectively. If from
this initial alignment
we randomly shuffle a selected portion of pairs of
entries in the inter-layer degree sequence $k_{out}$, we
decrease the value of $\xi$. In particular, if we randomly mix
all entries, then $k_{out}$
corresponds to a random permutation
of $k_{in}$ and the correlation term reads $\xi = \langle k \rangle^2$.
Finally, if the entries 
of $k_{in}$ are ordered in ascending (descending) order, 
while those of $k_{out}$ are sorted in descending (ascending) order,
the correlation term reaches the minimum value $\xi_{min}$~\cite{Dar73}.
When $\xi$ is in the range 
$[\langle k \rangle^2, \langle k^2\rangle]$,
the term in the square root of 
Eq.~(\ref{eq:thre}) is strictly
larger than zero: all regimes exist, and the corresponding 
phase transitions are discontinuous (see Fig.~2).
At the same time, the phase diagram suggests
that a further decrement of $\xi$ leads
to a drastic mutation in the features of the system.
For lower values of $\xi$, the analytic continuation
of our predictions tell us: (i) phase transitions
change their nature and become continuous; (ii)
the $\mathcal{I}$-phase is not longer present, and leaves
space to an hybrid $\mathcal{BD}$-phase where 
the system is simultaneously in both the bipartite
and decoupled regimes. The presence of this hybrid
phase appears evident by looking at the spectral properties
of the normalized laplacian even of finite size systems:
for $\xi=\xi_{min}$ and $p=0.5$, the order parameters
associated to the $\mathcal{B}$- and $\mathcal{D}$-phases
are in fact simultaneously larger than zero 
(see Figs.~2 and~S2).

\begin{figure}[!htb]
\begin{center}
\includegraphics[width=0.45\textwidth]{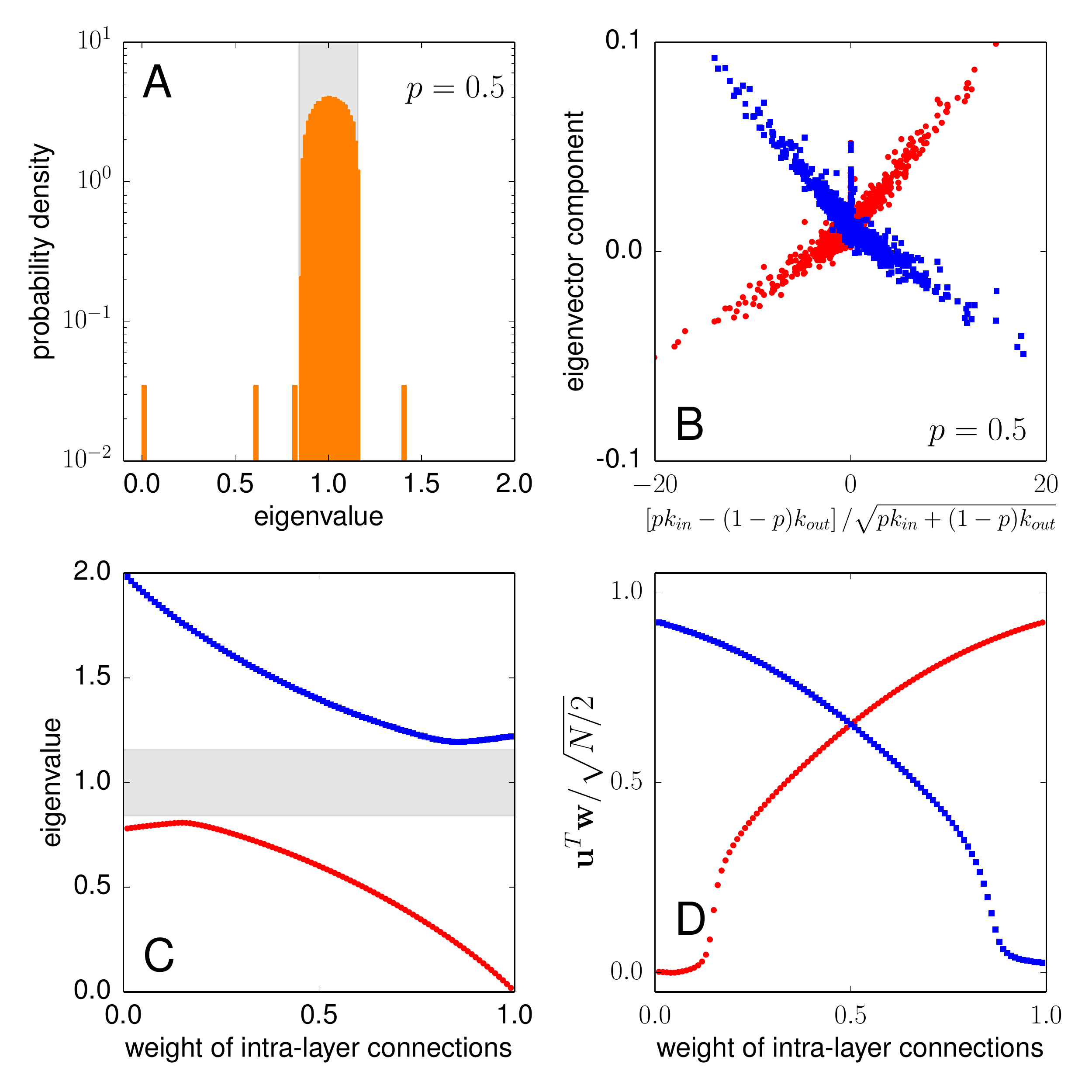}
\end{center}
\caption{
{\bf Supercritical regime}.
Spectral analysis of the normalized laplacian $\mathcal{L}$
for an interconnected network with 
degree sequence identical to the one analyzed in Fig.~1. 
In this case, however, intra- and inter-layer degree
sequences are aligned in such a way that the correlation
term is $\xi \simeq 1/2 (\langle k^2 \rangle + \langle k \rangle^2)$.
{\bf A} For $p=0.5$, the system is in the $\mathcal{BD}$-phase, 
where both the eigenvalues $\nu_2$ and
$\nu_{2N}$ are well separated from the
other eigenvalues of $\mathcal{L}$. {\bf B} The components
of both eigenvectors $\mathbf{w}_2$ and $\mathbf{w}_{2N}$
are correlated  with the r.h.s. of
Eq.~(\ref{eq:ans}). {\bf C} The eigenvalues $\nu_2$ (red circles)
and $\nu_{2N}$ (blue squares) touch the continuous band
tangentially.
{\bf D} The order parameter
clearly shows the presence of supercritical regime,
where both the $\mathcal{B}$- and $\mathcal{D}$-phases
are simultaneously present, for a wide range of values
of $p$. Only for very low (large) values of $p$, the
system is in the pure $\mathcal{B}$-phase ($\mathcal{D}$-phase).
}
\end{figure}

\

\noindent The previous effect is amplified
when we consider networks whose degrees obey a power-law degree 
distribution $P(k) \sim k^{-\gamma}$, with $\gamma<3$. 
Differently from the case of poissonian networks in fact, 
the correlation term $\xi$ has a much larger range of variability. 
Although the ansatz of Eq.~(\ref{eq:ans}) breaks down
for scale-free networks, and thus the entire analytic approach
is not working, Eq.~(\ref{eq:thre}) still tells
us what we should expect to see: unless the correlation term $\xi$ is very
close to its maximal value $\langle k^2\rangle$,
the term in the square root is negative; this means
that the $\mathcal{BD}$-phase is reached already for
little variations from the perfect alignment
between the intra- and inter-layer sequences $k_{in}$ and $k_{out}$.
This is indeed verified in our numerical experiments
as shown in Figs.~3 and~S3. For $p=0.5$, the presence of the 
hybrid $\mathcal{BD}$-phase is testified
by the fact that the eigenvalues
$\nu_2$ and $\nu_{2N}$ are simultaneously 
well separated from the other eigenvalues of $\mathcal{L}$, and
the associated eigenvectors $\mathbf{w}_2$ and $\mathbf{w}_{2N}$ have components
that are very well correlated with the r.h.s of
Eq.~(\ref{eq:ans}).
The mixing between the two degree
sequences causes a mismatch between intra- and inter-layer
hubs which do not correspond to the same nodes
as in the case of perfectly aligned
degree sequences. From the structural point of view, 
the total disappearance
of the so-called indistinguishable phase
can be interpreted as the absence of a buffer of structural safety. 
Scale-free networks seem therefore to be 
constantly at risks of potentially catastrophic failures.
There are, however, also dramatic
dynamical consequences. 
While for high values of $\xi$ there is
a neat distinction between the
$\mathcal{B}$- and $\mathcal{D}$-phases
and a random walker moves in the network 
performing either intra- or inter-layer diffusion, 
in the $\mathcal{BD}$-phase instead 
both motions happen simultaneously (Fig.~4).
Such ability makes the walker faster in the exploration of the 
entire system (Fig.~S4).

\begin{figure}[!htb]
\begin{center}
\includegraphics[width=0.45\textwidth]{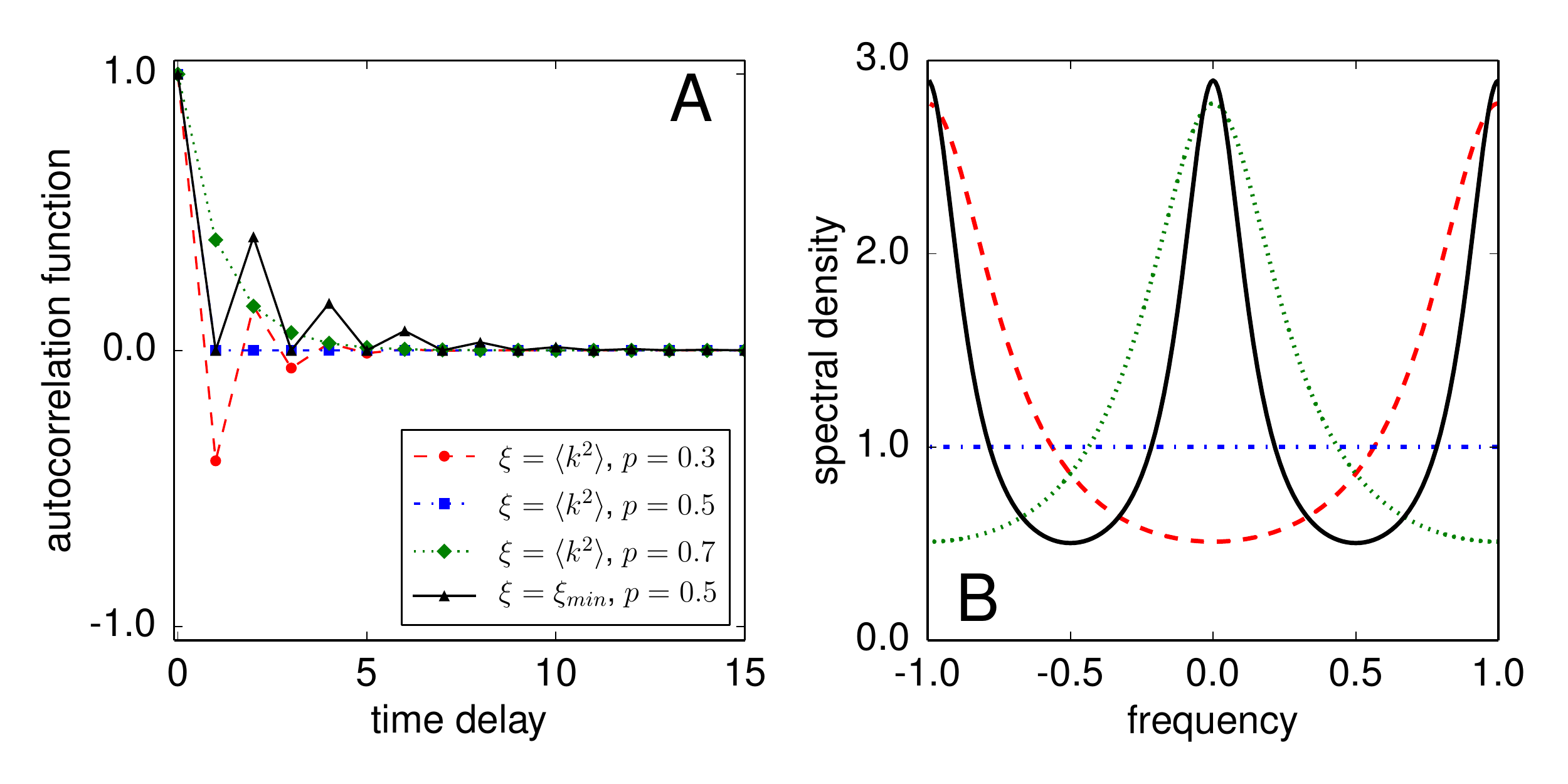}
\end{center}
\caption{
{\bf Random walk dynamics}.
We consider two interconnected networks composed of $N=8192$
nodes. Degrees are extracted from a power-law distribution
with exponent $\gamma=2.5$ defined on the support $[32,N]$. 
We monitor the trajectory of a random walker moving
on these systems for different values
of the correlation term $\xi$ and different
values of the weight of the intra-layer connections $p$. 
{\bf A} Autocorrelation function
$R(\tau) = \langle x(t) x(t+\tau) \rangle$
as a function of the time delay $\tau$. In the
definition of $R(\tau)$, $x(t)=+1$ if the random walker 
is on layer $A$ at time
$t$, and $x(t)=-1$, otherwise.
{\bf B} Spectral density $S(\omega)$ of the autocorrelation functions
of panel A. 
Intra-layer movements are characterized 
by the presence of a peak of $S(\omega)$ at
$\omega=0$, while inter-layer jumps
are emphasized by the presence of peaks
at $\omega=\pm 1$.
}
\end{figure}

\

\noindent To summarize, the
spectral properties of the normalized laplacian 
of random interconnected networks show that
these systems face a scenario that is much richer 
than the one valid for isolated networks. 
Depending on the combination of basic structural properties -- 
degree distribution, degree correlation, 
and strength difference between intra- and inter-layer connections -- 
topological and dynamical 
phase transitions associated
to the spectrum of the 
normalized laplacian can be either discontinuous or continuous, 
and different regimes can disappear or even coexist. This 
allows us, somehow, to divide the space of structural parameters 
in regions where the ``classical'' theory of isolated network holds, 
and those in which interdependent networks behave in anomalous manner.
As a final remark, we would like to stress
an appealing analogy between our
findings and the typical thermodynamic
behavior of any substance, e.g., water, carbon dioxide and methane.
In normal conditions, a substance changes
its phase from liquid to gas in a discontinuous manner.
However, above its critical temperature and pressure, the
substance ceases to exhibit distinct liquid- or 
gas-like state,
and it becomes a so-called supercritical fluid~\cite{McHugh86}. 
Since their properties can be fine-tuned, supercritical fluids have many
industrial and scientific applications. 
We believe that the possibility to drive
interconnected networks to supercritical regimes
could also be used in a fruitful way to better design and control
real world networked systems.

\begin{acknowledgements}
\noindent The author thanks A.~Arenas, C.~Castellano
and A.~Flammini for discussions and comments on 
the subject of this article.
\end{acknowledgements}

\bibliography{biblio}{}

\newpage

\onecolumngrid

\setcounter{figure}{0}
\setcounter{equation}{0}
\renewcommand{\thesection}{S \arabic{section}}
\renewcommand{\theequation}{S\arabic{equation}}
\renewcommand{\thefigure}{S\arabic{figure}}
\renewcommand{\thetable}{S\arabic{table}}

\section*{Supplementary Information}

\subsection*{Preliminary definitions}
\noindent We consider a symmetric and 
weighted interconnected network formed by two network
layers of identical size $N$.
The adjacency matrix $G$ of the entire network
can be written in the block form
\begin{equation}
G = 
\left(
\begin{array}{cc}
A & C
\\ 
C^T & B 
\end{array}
\right)
\; .
\label{eq:adj}
\end{equation}
$A$, $B$ and $C$ are
$N \times N$ square symmetric and weighted matrices.
$A$ and $B$ represent the weighted adjacency matrices
of the two network layers and thus contain information
only about intra-layer connections,
the matrix $C$ 
lists instead all the inter-layer connections
among nodes of different layers.
When comparing Eq.~(\ref{eq:adj}) with Eq.~(1) of the main text, please
consider that the factors $p$ and $1-p$ have been implicitly
included in the definition of Eq.~(\ref{eq:adj}).
We can reduce Eq.~(\ref{eq:adj})
Eq.~(1) of the main text if we perform the substitutions
$A \to p A$, $B \to p B$, $C \to (1-p)C$ and $C^T \to (1-p)C^T$.

\

\noindent For simplicity of notation, let us define the in-strength vector 
$\left|s^{in}_{A}\right> = A \left|1\right>$ of layer $A$
as the vector whose $i$-th component is given by
the sum the weights of all edges that connect node $i \in A$ 
to other nodes of network layer $A$.  
Similar definitions apply
also to the out-strength vector 
$\left|s^{out}_{A}\right> = C \left|1\right>$, and
to the analogues for layer $B$, that are
$\left|s^{in}_{B}\right> = B \left|1\right>$ and
$\left|s^{out}_{B}\right> = C^T \left|1\right>$.
Please note
that we are making use of the
standard bra-ket notation for vectors, and
we have indicated with $\left|1\right>$ the vector whose
components are all equal to one.
Clearly, the strength vectors of the network layers
$A$ and $B$ are simply given by
$\left|s_{A}\right> = \left|s^{in}_{A}\right> + \left|s^{out}_{A}\right>$
and $\left|s_{B}\right> = \left|s^{in}_{B}\right> + \left|s^{out}_{B}\right>$, respectively, while
the components of the vectors $\left|\Delta s_{A}\right> = \left|s^{in}_{A}\right> - \left|s^{out}_{A}\right>$ and $\left|\Delta s_{B}\right> = \left|s^{in}_{B}\right> - \left|s^{out}_{B}\right>$
quantify the difference between intra- and inter-layer 
connections of the various nodes.
Please note that all the former
vectors are composed of $N$ entries.

\subsection*{Normalized Laplacian}

\noindent Consider now the normalized laplacian matrix derived
from the adjacency matrix of Eq.~(\ref{eq:adj}), given by
\begin{equation}
\mathcal{L} = \mathbbm{1} - D^{-1/2} G D^{-1/2} \; ,
\label{eq:norm}
\end{equation}
where
\begin{equation}
D = \left(
\begin{array}{cc}
D_{A} & \emptyset \\
\emptyset & D_{B}
\end{array}
\right) \;.
\label{eq:diag}
\end{equation}
$D_{A}$ and $D_{B}$ are two diagonal matrices
such that the $i$-th diagonal elements 
are $\left(D_{A}\right)_{i,i} = \left(\left|s_{A}\right>\right)_i$ 
(i.e., the $i$-th component of the vector $\left| s_A \right>$) 
and $\left(D_{B}\right)_{i,i} = \left(\left|s_{B}\right>\right)_i$
(i.e., the $i$-th component of the vector $\left| s_B \right>$).

\

\noindent In order to solve in a simpler
way the eigenvalue problem 
\[
\mathcal{L} \left| w \right> = \nu \left|w\right> \, 
\] 
it is better to consider the matrix
\begin{equation}
D^{-1/2} \mathcal{L} D^{1/2} =  \mathbbm{1} - D^{-1} G \; ,
\label{eq:randw}
\end{equation}
where $\mathbbm{1}$ is the identity matrix, and $D^{-1}G$ is the
so-called random walk matrix.
Since $\mathcal{L}$ and $\mathbbm{1} - D^{-1} G$ are related by
a similarity transformation, they share the
same spectrum. The right eigenvector $\left|w\right>$ of $\mathcal{L}$
is related to the left eigenvector of
$\left|u\right>$  of $\mathbbm{1} - D^{-1} G$ by
\[
\left| w \right> = D^{-1/2} \left| u \right> \; ,
\]
while the right eigenvector $\left|w\right>$ of $\mathcal{L}$
is related to the right eigenvector of
$\left|v\right>$  of $\mathbbm{1} - D^{-1} G$ by
\[
\left| w \right> = D^{1/2} \left| v \right> \; ,
\]
and consequently the left eigenvector $\left|u\right>$ 
of $\mathbbm{1} - D^{-1} G$ and the right eigenvector of
$\left| v \right>$ of $\mathbbm{1} - D^{-1} G$ are related by
\[
\left|v\right> = D^{-1} \left|u\right> \; .
\]
For our purposes, it is even more convenient to rewrite the eigenvalue problem as follows
\[
\mathbbm{1} \left|v\right> - D^{-1} G \left|v\right> = \nu \left|v\right> \; ,
\]
\[
D^{-1} G \left|v\right> = (1-\nu) \left|v\right>
\]
and 
\[
G \left|v\right> = (1-\nu) D \left|v\right> \; .
\]
Please note that previous equation
has one trivial solution given by
given by $1-\nu =0$ and $\left|v\right> =\left|1\right>$, thus
any other eigenvector with different eigenvalue 
must be orthogonal to $\left|1\right>$.

\

\noindent If we write the eigenvector
$\left|v\right> = \left|v_A, v_B\right>$ as the composition
of two vectors $\left|v_A\right>$ and $\left|v_B\right>$ with $N$ entries each
corresponding to the nodes of layers $A$ and $B$, respectively, 
the previous equation is equivalent to the coupled equations
\begin{equation}
\begin{array}{l}
A \left|v_A\right> + C \left|v_B\right> = \left(1 -\nu\right) D_A \left|v_A\right>
\\
B \left|v_B\right> + C^T \left|v_A\right> = \left(1 - \nu\right) D_B \left|v_B\right>
\end{array} \;.
\label{eq:first}
\end{equation}

\noindent If we multiply these equations for $\left<1\right|$, we have
\[
\begin{array}{l}
 \left<1\right|A \left|v_A\right> + \left<1\right|C \left|v_B\right> = \left(1 - \nu\right) \left<1\right|D_A \left|v_A\right>
\\
\left<1\right|B \left|v_B\right> + \left<1\right|C^T \left|v_A\right> = \left(1 - \nu\right) \left<1\right|D_B \left|v_B\right>
\end{array}
\]
\[
\begin{array}{l}
 \left<s_A^{in}|v_A\right> + \left<s_B^{out}|v_B\right> = \left(1 - \nu\right) \left<s_A|v_A\right>
\\
\left<s_B^{in}|v_B\right> + \left<s_A^{out}|v_A\right> = \left(1 - \nu\right) \left<s_B|v_B\right>
\end{array}
\]
If we take
their difference, we have
\[
\left(1 - \nu\right) \left( \left<s_A|v_A\right> -  \left<s_B|v_B\right> \right)
= \left<\Delta s_A | v_A\right> - \left<\Delta s_B | v_B\right>
\]
if we finally suppose that $\left<s_A|v_A\right> - \left<s_B|v_B\right> \neq 0$, this becomes
\[
\nu = 1 - \frac{\left<\Delta s_A | v_A\right> - \left<\Delta s_B | v_B\right>}{\left<s_A|v_A\right> -  \left<s_B|v_B\right>} \; .
\]

\

\begin{figure}[!htb]
\includegraphics[width=0.75\textwidth]{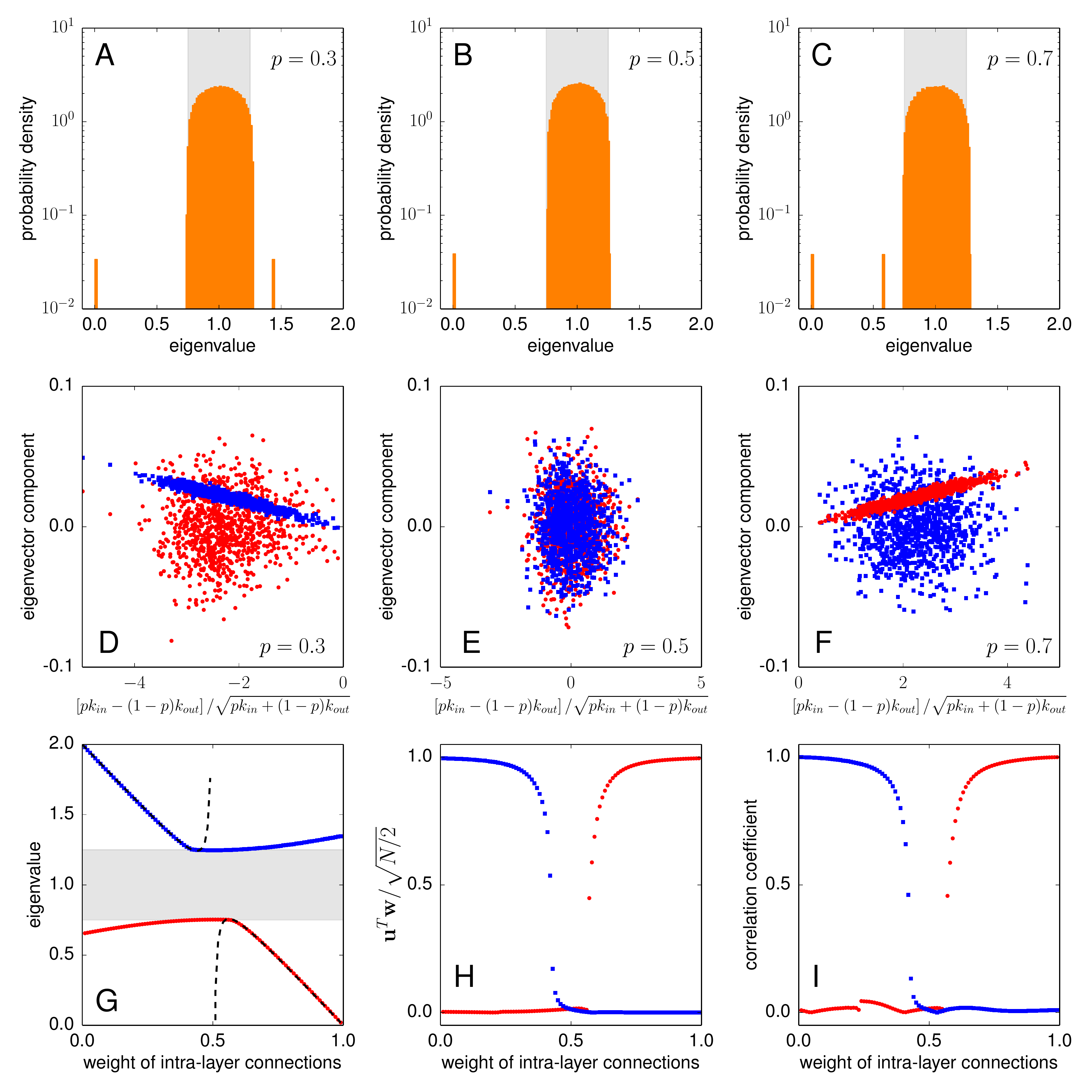}
\caption{
Spectral analysis of the normalized laplacian $\mathcal{L}$
of an interconnected network composed of two network layers 
of size $N=1024$ with intra- and inter-layer degrees obeying a 
Poisson distribution with average degree $\langle k \rangle =32$.
In- and out-degrees sequences
are randomly mixed so that the correlation term
reads $\xi = \langle k \rangle^2$.
The description of the various panels is identical to
the one of Fig.~1 of the main text. In panels D, E and F, the
components of the eigenvectors $\mathbf{w}_2$ and
$\mathbf{w}_{2N}$ are plotted as functions of the components
of the vectors appearing on 
the r.h.s. of Eq.~(\ref{eq:mod_ans}). In Panel G, 
the dashed line
stands for Eq.~(\ref{eq:norm3}).
}
\end{figure}

\noindent Please note that the graphical closure of
the graph imposes that 
\[
\left<1|s_A^{out}\right> = \left<1|s_B^{out}\right> \; .
\]
When the internal
strength of the both layers is comparable, i.e., 
\[
\left<1|s_A^{in}\right> \simeq \left<1|s_B^{in}\right> \; ,
\]
then numerical computations of the
spectrum of $\mathcal{L}$, such as
those presented in Fig.~S1, show that, the second smallest
eigenvector (largest eigenvector) in the
decoupled regime (bipartite regime) of the normalized laplacian
is approximately given by
\begin{equation}
\begin{array}{l}
\left|w_A\right > = n_A D_A^{-1/2} \left|\Delta s_A\right> \\
\left|w_B\right > = n_B D_B^{-1/2} \left|\Delta s_B\right> 
\end{array} \; ,
\label{eq:mod_ans}
\end{equation}
where $n_A$ and $n_B$ are normalization constants.
This means that
\begin{equation}
\begin{array}{l}
\left|v_A\right > = n_A \left|\Delta s_A\right> \\
\left|v_B\right > = n_B \left|\Delta s_B\right> 
\end{array} \; ,
\label{eq:mod_ans2}
\end{equation}
\noindent Since $\left|v_A, v_B\right>$ is
orthogonal to $\left|1,1\right>$, we can write
\[
n_A \left<1|\Delta s_A\right> + n_B \left<1|\Delta s_B\right> = 0 \; \leftrightarrow \; n_A = -n_B \frac{\left<1|\Delta s_B\right>}{\left<1|\Delta s_A\right>}
\]
thus
\[
\left<s_A|v_A\right> - \left<s_B|v_B\right> = -n_B \frac{\left<1|\Delta s_B\right>}{\left<1|\Delta s_A\right>} \left<s_A|\Delta s_A\right> - n_B \left<s_B|\Delta s_B\right> \; ,
\]
and 
\[
\begin{array}{l}
\left<\Delta s_A|v_A\right> - \left<\Delta s_B|v_B\right> = 
\\
-n_B \frac{\left<1|\Delta s_B\right>}{\left<1|\Delta s_A\right>} \left<\Delta s_A|\Delta s_A\right> - n_B \left<\Delta s_B|\Delta s_B\right>
\end{array} \; .
\]
So in conclusion, the eigenvalue corresponding
to this eigenvector is given by
\begin{equation}
\nu^* = 1 - \frac{\left<\Delta s_A | \Delta s_A\right> / \left<1|\Delta s_A\right>  + \left<\Delta s_B | \Delta s_B\right> / \left<1|\Delta s_B\right>}{\left<s_A|\Delta s_A\right> / \left<1|\Delta s_A\right> +  \left<s_B|\Delta s_B\right> / \left<1|\Delta s_B\right>} \; .
\label{eq:norm1}
\end{equation}
Please note that
\[
\left<\Delta s_A | \Delta s_A\right> = \sum_{i \in A} (s_i^{in} - s_i^{out}) (s_i^{in} - s_i^{out}) = \sum_{i \in A} (s_i^{in})^2 + (s_i^{out})^2 - 2 s_i^{in} s_i^{out}
= N \left(m_{2,A}^{in} + m_{2,A}^{out} - 2 \xi_A \right) \;,
\]
where $m_{2,A}^{in}$ is the second moment of the
in-strength distribution of layer $A$, 
$m_{2,A}^{out}$ is the second moment of the
out-strength distribution of layer $A$, and
$\xi_A$ is the correlation between the in- and out-strength
of nodes in layer $A$. Similarly, one
can write
\[
\left<\Delta s_A | s_A\right> = \sum_{i \in A} (s_i^{in} - s_i^{out}) (s_i^{in} + s_i^{out}) = \sum_{i \in A} (s_i^{in})^2 - (s_i^{out})^2 
= N \left(m_{2,A}^{in} - m_{2,A}^{out} \right) \; ,
\]
\[
\left<\Delta s_A | 1\right> = \sum_{i \in A} (s_i^{in} - s_i^{out})  
= N \left(m_{1,A}^{in} - m_{1,A}^{out} \right) 
\]
and
\[
\left<s_A | 1\right> = \sum_{i \in A} (s_i^{in} + s_i^{out})  
= N \left(m_{1,A}^{in} + m_{1,A}^{out} \right) \;,
\]
where $m_{1,A}^{in}$ and $ m_{1,A}^{out}$ are the first
moments of the in- and out-strength distribution, respectively.
The same relations are also valid for layer $B$.
Eq.~(\ref{eq:norm1})
can be thus rewritten as
\[
\nu^* = 1 - \frac{(m_{2,A}^{in} + m_{2,A}^{out} - 2 \xi_A ) / (m_{1,A}^{in} - m_{1,A}^{out} )  + (m_{2,B}^{in} + m_{2,B}^{out} - 2 \xi_B ) / (m_{1,B}^{in} - m_{1,B}^{out} )}{(m_{2,A}^{in} - m_{2,A}^{out} ) / (m_{1,A}^{in} - m_{1,A}^{out} )+ (m_{2,B}^{in} - m_{2,B}^{out} ) / (m_{1,B}^{in} - m_{1,B}^{out} )} \; .
\label{eq:norm2}
\]
If we finally assume that both network layers
have identical in- and out-strength distributions
and identical correlations between in- and out-strengths, the
former expression further simplifies to
\begin{equation}
\nu^* = 1 - \frac{m_{2}^{in}  +  m_{2}^{out} - 2\xi}{m_{2}^{in}  -  m_{2}^{out} } \; .
\label{eq:norm3}
\end{equation}
Please note that if we consider
the adjacency matrix of Eq.~1 of the main text,
we have to perform the following substitutions: 
$m_{2}^{in} \to p^2 \langle k^2\rangle$, 
$m_{2}^{out} \to (1-p)^2 \langle k^2\rangle$ and 
$\xi \to p (1-p)\xi$.
We thus recover Eq.~(7) of the main text.

\subsection*{Critical points}
The critical points of the transitions are calculated as
the values of $p$ for which Eq.~(7) intersects
the spectral band of Eq.~(4) predicted by Chung et al.
For $p \neq 1/2$, this leads to the equation
\[
\frac{2p^2 -2p +1}{2p -1} - 2 \frac{\xi}{ \langle k^2 \rangle} \frac{p (1-p)}{2p-1} = \pm  \frac{2}{\sqrt{ 2 \langle k \rangle}} \; ,
\]
from which
\[
(2p^2 -2p +1) \langle k^2 \rangle - 2 p (1-p) \xi \mp \frac{2}{\sqrt{ 2 \langle k \rangle}} \langle k^2 \rangle (2p -1) = 0
\]
and finally
\[
2p^2 \left( \langle k^2 \rangle + \xi \right) - 2 p \left[ \langle k^2 \rangle \left( 1 \pm  \frac{2}{\sqrt{2 \langle k \rangle}} \right) + \xi \right] +  \langle k^2 \rangle \left( 1 \pm  \frac{2}{\sqrt{2 \langle k \rangle}} \right) = 0 \; .
\]
Such quadratic equation has discriminant given by
\begin{equation}
\Delta = 4 \left[ \xi^2 +  \langle k^2 \rangle^2 \left(\frac{2}{\langle k \rangle} - 1 \right) \right] \; ,
\label{eq:discr}
\end{equation}
and thus solution given by
\[
p_c = \frac{1}{2} \left( 1 \pm \frac{ \frac{2 \langle k^2 \rangle} {\sqrt{2 \langle k \rangle}}  \pm  \sqrt{\xi^2 + \langle k^2 \rangle^2 \left(\frac{2}{\langle k \rangle} - 1 \right)} } {2 \left(\langle k^2 \rangle + \xi \right)} \right) \; .
\]
The solutions of this equations closer to $p=1/2$ are not
not interesting for us because $\nu^*$ has already entered
the spectrum at point. The effective critical points are
thus given by Eq.~(9) of the main text.

\subsubsection*{Positively/neutrally correlated network layers}
\noindent 
A simple strategy to explore this regime
of correlation is the following. We sort both the intra-layer
and inter-layer degree sequences in ascending (descending)
order, but we then shuffle, with probability $q$, 
each entry of the sequence with another 
randomly chosen entry. For $q=0$, the two sequences
are identical and the correlation term reads $\xi = \langle k^2 \rangle$.
In this case, we say that network layers are positively
correlated. For $q=1$, the out-degree sequence 
is effectively a random permutation
of $k_{in}$, and the correlation term reads $\xi = \langle k \rangle^2$.
We thus say that network layers are neutrally correlated.
For general values of $q$, the correlation term obeys
\[
\xi = (1-q) \langle k^2 \rangle + q \langle k \rangle^2 \; ,
\]
with $q \in [0,1]$.

\

\noindent The discriminant of Eq.~(\ref{eq:discr}) becomes
\[
\Delta = \left[(1-q) \langle k^2 \rangle + q \langle k \rangle^2 \right]^2 + \langle k^2 \rangle^2 \left(\frac{2}{\langle k \rangle} - 1 \right) \; ,
\]
, and thus
\[
\Delta = \left(1-2q+q^2+\frac{2}{\langle k \rangle} - 1\right) \langle k^2 \rangle^2 + 2 q (1-q)  \langle k^2 \rangle  \langle k \rangle^2 + q^2 \langle k \rangle^4 \; ,
\]
\[
\Delta =
\left(-2q+q^2+\frac{2}{\langle k \rangle}\right) \langle k^2 \rangle^2 + 2 q (1-q)  \langle k^2 \rangle  \langle k \rangle^2 + q^2 \langle k \rangle^4 \; ,
\]
\[
\Delta =
q^2 \left(  \langle k^2 \rangle^2 - 2  \langle k^2 \rangle  \langle k \rangle^2 +  \langle k \rangle^4 \right) -2q \langle k^2 \rangle \left( \langle k^2 \rangle - \langle k \rangle^2 \right) + \frac{2}{\langle k \rangle} \langle k^2 \rangle^2 \;, 
\]
and 
\[
\Delta =
q^2 \left(  \langle k^2 \rangle - \langle k \rangle^2 \right)^2 -2q \langle k^2 \rangle \left( \langle k^2 \rangle - \langle k \rangle^2 \right) + \frac{2}{\langle k \rangle} \langle k^2 \rangle^2 \; .
\]
The value $q_c$ of $q$ for which $\Delta = 0$ is therefore the solution of
a quadratic equation in $q$. This is given by
\[
q_c = \frac{1}{2 \left(  \langle k^2 \rangle - \langle k \rangle^2 \right)^2} \left[ 2 \langle k^2 \rangle \left( \langle k^2 \rangle - \langle k \rangle^2 \right) \pm \sqrt{ 4 \langle k^2 \rangle^2 \left( \langle k^2 \rangle - \langle k \rangle^2 \right)^2  -  4  \langle k^2 \rangle^2 \, 2/\langle k \rangle\,  \left(  \langle k^2 \rangle - \langle k \rangle^2 \right)^2  }   \right] \; ,
\]
from which we finally find
\[
q_c = \frac{\langle k^2 \rangle}{\langle k^2 \rangle - \langle k \rangle^2} \left[ \; 1 \pm \sqrt{1 - \frac{2}{\langle k \rangle}} \; \right] \; .
\]
Please note that $\frac{\langle k^2 \rangle}{\langle k^2 \rangle - \langle k \rangle^2} \geq 1$, thus
the only solution that is potentially
smaller or equal to one
\[
q_c = \frac{\langle k^2 \rangle}{\langle k^2 \rangle - \langle k \rangle^2} \left[ \; 1 - \sqrt{1 - \frac{2}{\langle k \rangle}} \; \right] \; .
\]
For poissonian graphs, the former expression shows
that $q_c > 1$. Thus for any value of $q \in [0,1]$
$\Delta > 0$. For scale-free graphs with degree exponent
$\gamma <3$ instead, 
$\frac{\langle k^2 \rangle}{\langle k^2 \rangle - \langle k \rangle^2} \simeq 1$, and $q_c \leq 1$.

\subsubsection*{Negatively/neutrally correlated network layers}
\noindent To explore this regime
of correlation, we sort both the intra-layer in ascending (descending)
order and inter-layer degree sequences in descending (ascending)
order, but we then shuffle, with probability $t$, 
each entry of the sequence with another 
randomly chosen entry. For $t=0$, the correlation term reads $\xi = \xi_{min}$
and the layers are negatively correlated.
For $t=1$, the out-degree sequence is effectively a random permutation
of $k_{in}$, and the correlation term reads $\xi = \langle k \rangle^2$.
Network layers are neutrally correlated.
For general values of $t$, the correlation term obeys
\[
\xi = (1-t) \xi_{min} + t \langle k \rangle^2 \; ,
\]
with $t \in [0,1]$. 
The discriminant is therefore
\[
\Delta = \left[(1-t) \xi_{min} + t \langle k \rangle^2 \right]^2 + \langle k^2 \rangle^2 \left(\frac{2}{\langle k \rangle} - 1 \right) \; .
\]
This becomes
\[
\Delta = t^2 \left(\xi_{min} - \langle k \rangle^2\right)^2 -2t \xi_{min}  \left(\xi_{min} -  \langle k \rangle^2 \right) + \xi_{min}^2 +  \langle k^2 \rangle^2 \left(\frac{2}{\langle k \rangle} - 1 \right) \; .
\]
Thus the critical value of $t$ is given by
\[
t_c =  \frac{1}{\xi_{min} - \langle k \rangle^2} \left[ \; \xi_{min} \pm \langle k^2 \rangle \sqrt{1 -  \frac{2}{\langle k \rangle}} \; \right] \; .
\]
The solution with the plus sign is always negative, thus
\[
t_c =  \frac{\langle k^2 \rangle \sqrt{1 -  \frac{2}{\langle k \rangle}} - \xi_{min}}{\langle k \rangle^2-\xi_{min}}  \; .
\]

\begin{figure}[!htb]
\includegraphics[width=0.75\textwidth]{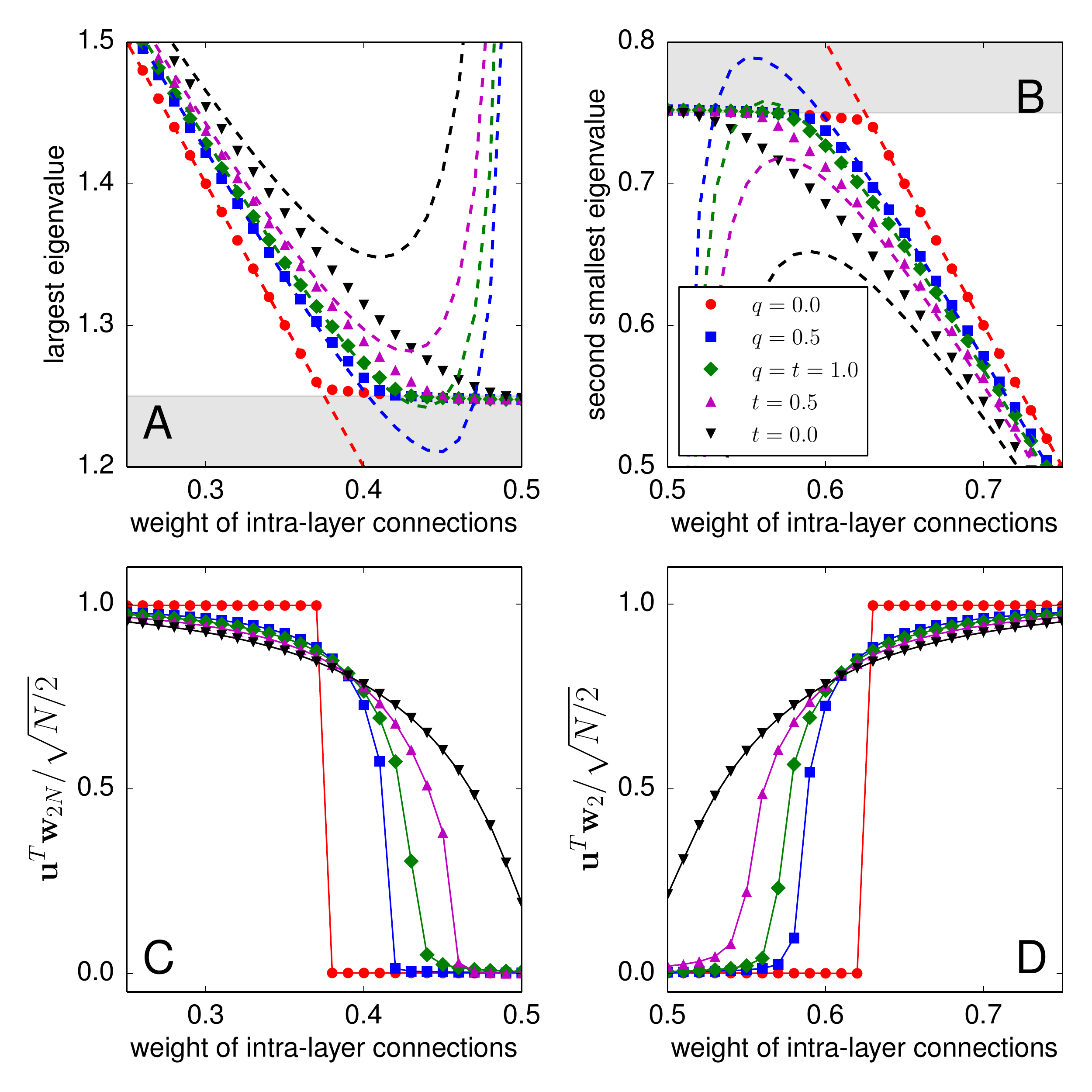}
\caption{
We consider interconnected networks
composed of two network layers 
of size $N=8192$ with intra- and inter-layer degrees obeying a 
Poisson distribution with average degree $\langle k \rangle =32$. 
As explained in the text, the continuous parameters $q$ and $t$
are used to control
the alignment between the entries of the two
degree sequences sequences, so that
the correlation term reads 
$\xi = (1-q)\langle k^2 \rangle + q \langle k\rangle^2$
or $\xi = (1-t)\xi_{min} + t \langle k\rangle^2$.
{\bf A} For any value of
$q$, the largest eigenvalue $\nu_{2N}$
intersects the continuous band (gray area). 
The corresponding transition, in the
thermodynamic limit, is expected
to be discontinuous. The results of numerical computations 
(symbols) are in very
good agreement with our analytic predictions (dashed lines).
For $t<1$, our predictions
fail to correctly describe
how $\nu_{2N}$ changes as a function of the
weight of the intra-layer connections.  
{\bf B} Same as for panel A, but for the second
smallest eigenvalue
$\nu_{2}$. {\bf C} Sum of the
components of $\mathbf{w}_{2N}$ corresponding
to nodes of a single layer.
{\bf D} Same as for panel C, but for the second smallest 
eigenvector $\mathbf{w}_{2}$.
}
\end{figure}

\begin{figure}[!htb]
\includegraphics[width=0.75\textwidth]{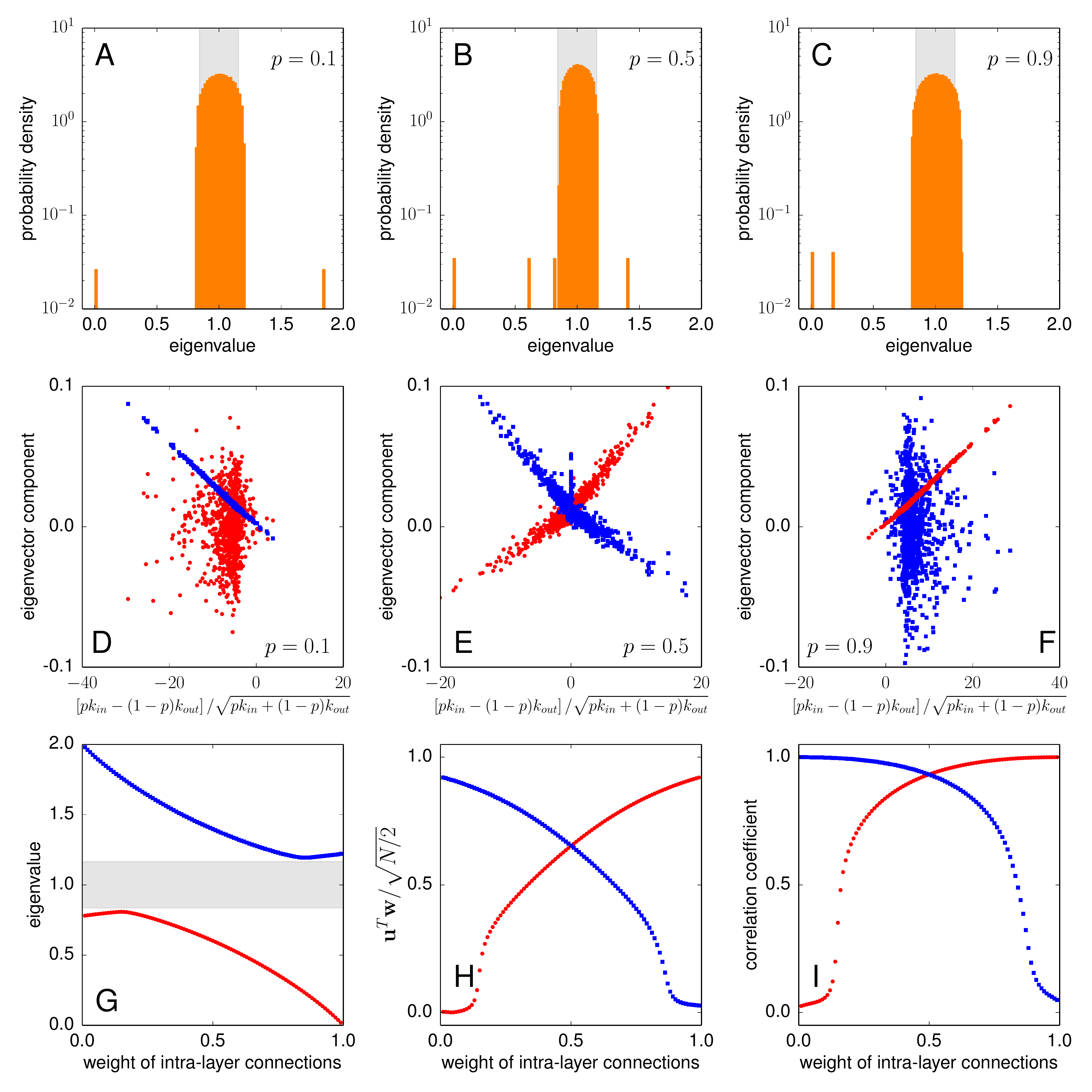}
\caption{
Spectral analysis of the normalized laplacian $\mathcal{L}$
the scale-free interconnected network analyzed in Fig.~3
of the main text. 
{\bf A} When the weight of the intra-layer connections $p$ is
sufficiently low, the system is in a pure bipartite regime. 
{\bf B} After that, the system enters in a hybrid
phase, where both eigenvalues $\nu_2$ and $\nu_{2N}$ are
simultaneously outside the continuous band. {\bf C} For large values
of $p$, $\nu_{2N}$ gets absorbed in the continuous band, 
and only $\nu_2$ is detached from the rest of the spectrum.
{\bf D} In the pure bipartite regime, the
components $\mathbf{w}_{2N}$  (blue squares) of a single
layer are correlated with the r.h.s. of
Eq.~(\ref{eq:mod_ans}), while those of 
$\mathbf{w}_{2}$  (red circles) are not.
{\bf E} In the hybrid regime, the components of both eigenvectors 
$\mathbf{w}_{2}$ and $\mathbf{w}_{2N}$  are
correlated with the r.h.s. of
Eq.~(\ref{eq:mod_ans}). {\bf F} In the pure decoupled regime, 
only the components of $\mathbf{w}_{2}$ 
are correlated with the r.h.s. of
Eq.~(\ref{eq:mod_ans}).
{\bf G} The eigenvalues $\nu_2$ (red circles)
and $\nu_{2N}$ (blue squares) touch the continuous band
tangentially. This is an indication of continuous transitions between
the various regimes. {\bf H} The sum of the eigenvector components
of a single network layer clearly show the coexistence
of the two dynamical phases.
{\bf I} The same is also visible if we monitor 
the state of the system with 
the correlation 
coefficient between
the components of the eigenvectors and the 
r.h.s. of Eq.~(\ref{eq:mod_ans}).
}
\end{figure}

\subsection*{Positively correlated layers}
This is a special case in which we can
determine an additional
eigenpair of $\mathcal{L}$. 
If $\left|s_A^{in}\right> = \left|s_B^{in}\right> = p \left|k\right>$,
and $\left|s_A^{out}\right> = \left|s_B^{out}\right> = (1-p) \left|k\right>$,
then
\[
D^{-1} G = 
\left(
\begin{array}{cc}
D_A^{-1} & \emptyset
\\
\emptyset & D_B^{-1}
\end{array}
\right)
\left(
\begin{array}{cc}
p A & (1-p)C
\\
(1-p)C^T & p B
\end{array}
\right)
=
\left(
\begin{array}{cc}
K^{-1} & \emptyset
\\
\emptyset & K^{-1}
\end{array}
\right)
\left(
\begin{array}{cc}
p A & (1-p)C
\\
(1-p)C^T & p B
\end{array}
\right)
\]

\[
\left(
\begin{array}{cc}
K^{-1} & \emptyset
\\
\emptyset & K^{-1}
\end{array}
\right)
\left(
\begin{array}{cc}
p A & (1-p)C
\\
(1-p)C^T & p B
\end{array}
\right)
\left(
\begin{array}{c}
\left|1\right>
\\
-\left|1\right>
\end{array}
\right)
=
\left(
\begin{array}{cc}
K^{-1} & \emptyset
\\
\emptyset & K^{-1}
\end{array}
\right)
\left(
\begin{array}{c}
2p-1 \left|k\right>
\\
1-2p\left|k\right>
\end{array}
\right)
=
(2p-1)
\left(
\begin{array}{c}
\left|1\right>
\\
-\left|1\right>
\end{array}
\right) \; .
\]

\noindent This means that the vector $\left|1,-1\right>$ is
eigenvector of $K^{-1}G$ with eigenvalue $(2p-1)$.
Thus the vector $K^{1/2} \left|1,-1\right>$ is
eigenvector of $\mathcal{L}$ with eigenvalue $2(1-p)$.
Please note that this eigenvector should be normalized
such that 
\[
1= n^2 \left<1,-1\right| K^{1/2}  K^{1/2} \left|1,-1\right> = n^2 \left<1,-1\right| K \left|1,-1\right> = n^2 \left<1,-1\right. \left|k,-k\right> = n^2 2 \left<1|k\right>
\]
and thus the normalization reads
\[
n = \frac{1}{\sqrt{2 \left<1|k\right> }} \; .
\]
In conclusion $\mathcal{L}$ has always associated
the eigenpair
\[
\nu^* = 2(1-p) \qquad \textrm{ and } \qquad \left|w^*_A, w^*_B\right> = \frac{1}{\sqrt{2 \left<1|k\right>}} K^{1/2} \left|1,-1\right> \; ,
\]
with $K$ diagonal matrix whose diagonal elements correspond to
the degree of the nodes.

\begin{figure}[!htb]
\includegraphics[width=0.45\textwidth]{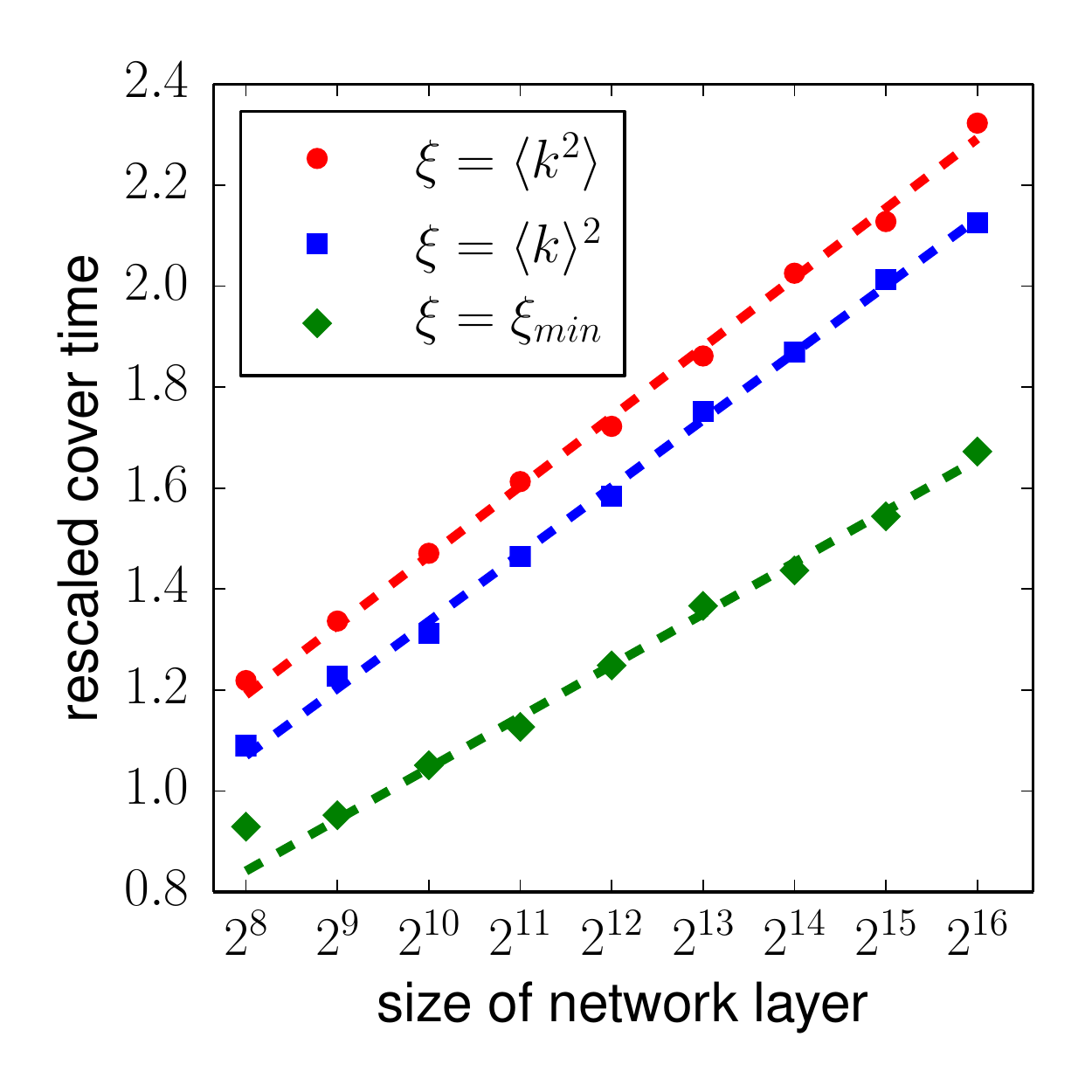}
\caption{
Cover time $T_{cover}$ of a random walker moving in an interconnected network composed
of two layers of $N$ nodes. The degrees are extracted from a 
power-law distribution with exponent $\gamma=2.5$ defined on the
support $[5,N]$. Cover time is divided by
the sum of the strengths of all nodes
in the network. The weight $p$ of intra-layer
connections is set equal to $0.5$. Each point 
represents the average value of the normalized cover time obtained in
at least $100$ realizations of the network models.
The dashed lines stand for fits with the function
$T_{cover} = a + b \log{N}$. We measure $b=0.20$ (red), $b=0.19$ (blue)
and $b=0.15$ (green).
}
\end{figure}

\subsection*{Autocorrelation function of random walk dynamics}

Indicate with $\left|\pi\right>$ the 
vector whose $i$-th component 
represents the stationary distribution
to find the random walker on node $i$, i.e., 
\[
\left|\pi\right>_i = \frac{\left|s\right>_i}{\left<1|s\right>} \; .
\]
Suppose now the walker starts its walk at position $j$.
The probability to find the walker at a given position after
$t$ steps is given contained in the vector
\[
\left< z^{(t)}_j \right| = \left<j\right|\left(D^{-1} G \right)^t \;,
\]
where we indicated with $\left|j\right>$ the vector
whose components are all equal to zero except for its
$j$-th component that equals one, and 
with $\left(D^{-1} G \right)^t$ the $t$-th power of the
transition matrix $D^{-1}G$.

\

Suppose we monitor
the position of the random walker
at time $\tau$ throught the variable $x(\tau)$.
We define $x(\tau) = +1$ if the walker is in one of the
nodes of layer $A$ at stage
$\tau$, and $x(\tau)=-1$, otherwise. The average
value of $x(\tau|j)$ given that the initial position is
on node $j$ is thus given by
\[
\langle x(\tau|j) \rangle =  \left< z^{(\tau)}_j | 1,-1 \right> = \left<j| (D^{-1} G)^{\tau} | 1,-1 \right> \; , 
\]
where the vector $\left|1,-1 \right>$ is a vector whose
first $N$ components are all equal to $1$, and its second $N$
components are equal to $-1$.
Since, at stationarity, the proability to start at position
$j$ is equal to $\left|\pi\right>_j$, we can write
\[
\langle x(\tau) \rangle =  \sum_j\;  \left<j| (D^{-1} G)^{\tau} | 1,-1 \right> \; \left|\pi\right>_j = \left<\pi| (D^{-1} G)^{\tau} | 1,-1 \right> \; ,
\]
as the average value of the variable $x(\tau)$ from an
arbitrary starting position. Finally, if we want to calculate
the autocorrelation function $R(\tau) = \langle x(0) x(\tau) \rangle$, we
need to insert in the former expression the value of $x(0|j)$. By
definition, we have $x(0|j) =1$ if $ j \in A$, and $x(0|j)=-1$ if $j \in B$.
We thus obtain
\begin{equation}
R(\tau) = \langle x(0) x(\tau) \rangle =  \sum_j\;  \left<j| (D^{-1} G)^{\tau} | 1,-1 \right> \; \left|\pi\right>_j x(0|j) = \left<\pi_A, - \pi_B| (D^{-1} G)^{\tau} | 1,-1 \right> \; ,
\label{eq:corr_si}
\end{equation}
where $\left|\pi_A\right>$ is the vector whose components
are equal to the stationary probabilities of nodes
within layer $A$, and $\left|- \pi_B \right> =- \left| \pi_B \right>$
is the vector whose components
are equal to the stationary probabilities of nodes
within layer $B$ multiplied by $-1$.

\subsubsection*{Positively correlated layers}
As shown before, if $\left|s_A^{in}\right> = \left|s_B^{in}\right> = p \left|k\right>$,
and $\left|s_A^{out}\right> = \left|s_B^{out}\right> = (1-p) \left|k\right>$,
then the vector $\left|1,-1\right>$ is a (non normalized) eigenvector of the matrix $D^{-1}G$ with
eigenvalue $\lambda = 2p-1$. If we use the eigendecomposition of $D^{-1}G$ and write
\[
\left(D^{-1} G\right)^\tau = \sum_{i} \lambda_i^\tau \,  \left|v_{i,A}, v_{i,B} \right>  \left<v_{i,A}, v_{i,B} \right|
\]
where $\left|v_{i,A}, v_{i,B} \right>$ is the $i$-th eigenvector of the matrix $D^{-1}G$, then we have
\begin{equation}
R(\tau) = \sum_{i} \lambda_i^\tau \,  \left<\pi_A, - \pi_B|v_{i,A}, v_{i,B} \right>  \left<v_{i,A}, v_{i,B} | 1,-1 \right> = 
(2p-1)^\tau  \frac{\left<\pi_A, - \pi_B|1, -1 \right>}{\sqrt{2N}} \, \frac{\left<1, -1 | 1,-1 \right>}{\sqrt{2N}} = (2p-1)^\tau \; .
\label{eq:corr_pos}
\end{equation}
Essentially, all the eigenvectors of $D^{-1}G$ are orthogonal to $\left|1,-1\right>$, and thus
$\left<v_{i,A}, v_{i,B} | 1,-1 \right> = 0$ for all $i$ unless $\left|v_{i,A}, v_{i,B} \right> = \frac{\left|1,-1\right>}{\sqrt{2N}}$.
We also used that $\left<\pi_A, -\pi_B|1,-1\right> = \left<\pi_A|1\right> + \left<\pi_B|1\right>=1$
and $\left<1,-1|1,-1\right> = 2N$.

\

Please note that in this special case it is
also very simple to calculate the probability
to stay on the same layer for $t_s$ consecutive steps.
Since the probability to stay on the same layer is
equal to $p$ for all nodes, we have
\[
P_{stay}(t_s) = p^{t_s} \; .
\]
In the same way, the probability to change
layer for $t_s$ consecutive steps
is
\[
P_{change}(t_s) = (1-p)^{t_s} \; .
\]

\

From the previous expression of $R(\tau)$, we
can also easily calculate the fourier transform
\[
F(\omega) = \sum_{\tau = 0}^\infty \, R(\tau) \, e^{-i \omega \tau} \;.
\]
Since $R(\tau) = (2p-1)^\tau$, we have
\[
F(\omega) = \sum_{\tau = 0}^\infty \, \left[(2p-1) \, e^{-i \omega} \right]^{\tau} =
\frac{1}{1- (2p-1) e^{-i \omega}} = \frac{[1 - (2p-1) \cos{\omega}] - i  (2p-1) \sin{\omega}} {[1 - (2p-1) \cos{\omega}]^2 +  \left[(2p-1) \sin{\omega}\right]^2 } \; .
\]
The spectral density is thus given by
\begin{equation}
S(\omega) = \frac{1}{4p^2-2(2p-1)(1+\cos{\omega})} \; .
\label{eq:spectral}
\end{equation}

\subsubsection*{General case}

In the general case, we can calculate the first values
of the autocorrelation function. For $\tau=0$, we
have
\[
R(\tau=0) = \left<\pi_A, - \pi_B | 1,-1 \right> = \left<\pi_A|1\right> + \left<\pi_B|1\right> =1 \;.
\]
For $\tau=1$, we have
\[
\begin{array}{l}
R(\tau=1) = \left<\pi_A, - \pi_B| D^{-1} G | 1,-1 \right> = \frac{1}{\left<s|1\right>} \, \left<1,-1|G|1,-1\right> = \frac{1}{\left<s_A|1\right> + \left<s_B|1\right>} \left<1,-1|\Delta s_A,-\Delta s_B\right> 
\\
= \frac{\left<1|\Delta s_A\right> + \left<1|\Delta s_B\right>}{\left<s_A|1\right> + \left<s_B|1\right>} = \frac{m^{in}_{1,A} - m^{out}_{1,A} + m^{in}_{1,B} - m^{out}_{1,B}}{m_{1,A} + m_{1,B}} 
\end{array}
\; .
\]
In particular, if $m^{in}_{1,A} = m^{in}_{1,B} = p \langle k \rangle$, 
$m^{out}_{1,A} = m^{out}_{1,B} = (1-p) \langle k \rangle$, and
$m_{1,A} = m_{1,B} = \langle k \rangle$, we have
\[
R(\tau=1) = (2p-1) \; .
\]


\end{document}